\documentclass[11pt,a4paper]{article}

\usepackage{amsmath}
\usepackage{amsfonts}
\usepackage{amssymb}

\usepackage{psfrag}
\usepackage{graphicx}

\newcommand{\vf}{\varphi}
\newcommand{\p}{\partial}
\newcommand{\nn}{\nonumber}
\newcommand{\ra}{\rightarrow}

\newcommand{\tr}{\tilde{r}}
\newcommand{\bvf}{\bar{\varphi}}
\newcommand{\dvf}{\delta\varphi}
\newcommand{\psih}{\psi_{\mathrm{hor.}}}
\newcommand{\dvfh}{\delta\varphi_{\mathrm{hor.}}}

\begin{document}
\title{\bf{Discreteness of dyonic dilaton black holes}}
\author{E. A. Davydov\thanks{davydov@theor.jinr.ru}\\
\small{\it{Bogoliubov Laboratory of Theoretical Physics, JINR, }}\\
\small{\it{6 Joliot-Curie St, Dubna, Moscow region, 141980,
Russian Federation,}}\\
\small{and}
\\
\small{\it{Peoples' Friendship University of Russia (RUDN
University),}}\\
\small{\it{6 Miklukho-Maklaya St, Moscow, 117198, Russian
Federation}}}

\maketitle

\begin{abstract}
We show that there are two classes of solutions that describe
static spherically symmetric dyonic dilaton black holes with two
nonsingular horizons. The first class includes only the already
known solutions that exist for a few special values of the dilaton
coupling constant. Solutions belonging to the second class have
essentially different properties. They exist for continuously
varying values of the dilaton coupling constant, but arise only
for discrete values of the dilaton field at the horizon. For each
given value of the dilaton coupling constant, there may exist
several such solutions differing by the number of zeros of the
shifted dilaton function in the subhorizon region and separating
the domains of singular solutions.

\end{abstract}

\section{Introduction}

Now, after decades of purely theoretical research, a new era
begins in the study of black holes. Thanks to such projects as the
Event Horizon Telescope, LIGO, VIRGO it will be possible to obtain
observational data and to test various theoretical models.
Therefore, it is important to close some gaps in the theories of
black holes that have not been resolved in due time. One of these
not fully investigated questions is the dyonic black hole model in
the presence of a dilaton field. Static spherically symmetric
black holes with electric and / or magnetic charges are well
studied, and solutions for them are represented by the
Reissner–-Nordstr\"{o}m metric. However, in the presence of the
dilaton field, the situation becomes much more complicated.

Dilatonic scalar field arises in a variety of theories including
dimensional reductions, low-energy limit of string theory, various
models of supergravity. A dilaton field naturally couples to gauge
fields and thus can influence various physical phenomena. Such a
dilaton effect is clearly manifested in models of black holes with
gauge fields. Now purely electrically or magnetically charged
black holes do not contain the RN solution as a special case. The
analytical solutions for single-charged dilatonic black hole with
any value of the dilaton coupling constant are
known~\cite{Gibbons:1987ps,Garfinkle:1990qj}, and in non-extremal
case they are all singular on the inner horizon (except for the
case of the zero dilaton coupling constant, which is trivial). If
the black hole carries both magnetic and electric charges it
admits the RN solution with a constant dilaton. Non-extremal
dyonic solutions also turn out to be singular in the general case
on one of the horizons, however, with the exception of some
special cases. The study of the conditions for the existence of
regular solutions (which do not have singularities on the
horizons) is a rather important problem. It was not brought to an
end so far, and we hope that this study will shed light on this
issue.

It should be mentioned that by now only two exact solutions for
the static spherically symmetric dyon-dilaton black hole are
known~\cite{Gibbons:1987ps,Dobiasch:1981vh,Gibbons:1982ih,Lee:1984hm}.
They arise for special values of the dilaton coupling constant:
$a=1$ and $a=\sqrt{3}$. These solutions are regular. In the study
of numerical solutions, it was found in~\cite{Poletti:1995yq} that
for values of the dilaton coupling constant close (or coincident)
to a sequence of triangular numbers,
\begin{equation}\label{A}
 a^2=n(n+1)/2\,,\quad n=0,1,2,\ldots,
\end{equation}
the corresponding solutions are also regular on both horizons.
Taking into account that the first two terms of this sequence
coincide with the values of $a$ corresponding to known exact
solutions, it was suggested that regular black hole solutions in
the dyon-dilaton system are possible only for these discrete
values of the dyon coupling constant. This conjecture we shall
call for brevity the \emph{triangular hypothesis}.

The first two values of the dilaton coupling from this triangular
sequence have a clear physical meaning: they arise in string
theory and with the Kaluza-Klein dimensional reduction. Attempts
were made to find physical theories in which the values of the
dilaton constant, corresponding to other members of this sequence,
may naturally appear. However, these attempts were
unsuccessful~\cite{Gibbons:1993xt,Nozawa:2010rf}. Therefore, the
triangular hypothesis was in a suspended state: without an
explicit physical justification, but with some purely mathematical
indications in its favor. For example, in the study of extreme
dyon black holes, it was found~\cite{Galtsov:2014wxl} that the
solution for the dilaton is analytic on the horizon only for
triangular values of the dilaton constant.

In this paper we verify the validity of the triangular hypothesis
and find out the conditions for the existence of regular solutions
for arbitrary values of the dilaton coupling constant in the model
of a static spherically-symmetric dyonic dilaton black hole. In
Section~II we give a description of the model and find the reasons
for the appearance of a triangular sequence~(\ref{A}). Since all
known analytical solutions in this model are associated with
solutions to Toda lattice equations, in Section~III we try to find
all configurations that can have a connection with Toda lattices
and construct the corresponding solutions. Section IV is devoted
to the investigation of numerical solutions for a much more
extensive domain in the parameter space than was done
in~\cite{Poletti:1995yq}. In the Conclusion we summarize and
discuss the results obtained.

\section{The Model}

Dyon system with dilaton field can be described by the following
action in $D\geq 4$ dimensions~\cite{Poletti:1995yq}:
\begin{equation}
    S=\int\!\left[R-\frac{D-2}{4(D-3)}(\nabla\vf)^2-\frac{e^{a\vf}}{2}\left(F_{[2]}^2+
    \frac{2}{(D-2)!}G_{[D-2]}^2\right)\right]\!\sqrt{-g}\,d^D x\,.\label{I1}
\end{equation}
The gauge part of the action~(\ref{I1}) contains the standard
Maxwell two-form $F_{\mu\nu}$ and the additional Abelian
$(D-2)$-form $G_{\mu_1..\mu_{D-2}}$. Gauge fields interact with
dilaton $\vf$ via the exponential potential characterized by the
coupling constant $a$. In comparison with the notations often
encountered in the literature, we multiplied the dilaton by a
constant factor $\frac{D-2}{4\sqrt{D-3}}$, depending on the number
of spacetime dimensions, $D$. This led to the appearance of an
unusual factor in the dilaton kinetic term, but many of the
further formulas will have a simpler form.

A static spherically symmetric metric can be written as:
\begin{equation}\label{I2}
    ds^2=-h(r)\,dt^2+h(r)\,s^2(r)\,dr^2+\rho^{2/(D-2)}\!(r)\,d\Omega^2_{D-2}\,,
\end{equation}
where $d\Omega_{D-2}^2=g_{MN}(y)\,dy^M dy^N$ is a metric on
$(D-2)$-dimensional sphere. The variable $s(r)$ represents the
remaining gauge degree of freedom which will be fixed later. The
curvature scalar density for such metric reads as:
\begin{equation}\label{I3}
    R\sqrt{-g}=\frac{h'\rho'}{hs}+(D-3)(D-2) hs\rho^{\frac{D-4}{D-2}}+\frac{(D-3)\rho'^2}{(D-2)\rho s}+\mbox{tot.
    div.}
\end{equation}

The equations of motion for gauge field are quite simple:
\begin{eqnarray}
  &&\nabla_{\mu}\left(e^{a\vf}F^{\mu\nu}\right) = 0\,, \\
  &&\nabla_{\mu_1}\left(e^{a\vf}G^{\,\mu_1..\mu_{D-2}}\right) = 0\,.
\end{eqnarray}
The appropriate field ansatz preserving spherical symmetry can be
written in the following form:
\begin{equation}\label{I4}
    F_{tr}=\frac{Q }{\rho}hs e^{-a\vf} \,\varepsilon_{tr}\,,\quad
    G_{M_1..M_{D-2}}=P\,\varepsilon_{M_1..M_{D-2}}\,,
\end{equation}
where $Q$ and $P$ are some constants, which can be called electric
and magnetic charges; $\varepsilon_{tr}$ and
$\varepsilon_{M_1..M_{D-2}}$ are antisymmetric tensors. We
consider only the dyonic configuration, for which both charges are
non-vanishing. Mention that in $D=4$ case, one can assume that
$F_{[2]}$ and $G_{[2]}$ are identical and describe the same
electromagnetic field.

With the ansatz~(\ref{I4}), the terms with gauge fields in the
action~(\ref{I1}) can be replaced by an effective potential. As a
result, the effective one-dimensional action for the dyon-dilaton
model~(\ref{I1}) can be written as
\begin{equation}
    S_{\mathrm{eff}}=\int\left(\frac{h'\rho'}{sh}+\frac{(D-3)\rho'^2}{(D-2)\rho s}-\frac{(D-2)\rho\vf'^2}{4(D-3)s}-hs\,U(\rho,\vf)\right)dr\,, \label{I6}
\end{equation}
where
\begin{equation}\label{I5}
    U(\rho,\vf)=\frac{1}{\rho}\left(P^2e^{a\vf}+Q^2e^{-a\vf}\right)-(D-3)(D-2) \rho^{\frac{D-4}{D-2}}\,.
\end{equation}

Varying it with respect to metric and dilaton variables, we obtain
the remaining equations of motion. The metric function $s$ is just
a Lagrange multiplier, and variation with respect to it provides
the Hamiltonian constraint:
\begin{equation}\label{I7}
    H=\frac{h'\rho'}{sh}+\frac{(D-3)\rho'^2}{(D-2)\rho s}-\frac{(D-2)\rho\vf'^2}{4(D-3)s}+hs\,U(\rho,\vf)=0\,.
\end{equation}
After that, one can choose the gauge as desired. The conventional
choice is Schwarzschild gauge $s=h^{-1}$, in which the equations
of motion for $\rho$, $h$ and $\vf$ read as:
\begin{eqnarray}
  &&h''+\frac{2(D-3)}{D-2}\left(\frac{h\rho'}{\rho}\right)' =-\frac{(D-3)h\rho'^2}{(D-2)\rho^2} -\frac{(D-2)h\vf'^2}{4(D-3)} -U_\rho\,,\label{S1}\\
  &&\rho''=\frac{(D-3)\rho'^2}{(D-2)\rho}-\frac{(D-2)\rho\,\vf'^2}{4(D-3)}\,, \label{Ss}\\
  &&(h\rho\, \vf')'=\frac{2(D-3)}{D-2}\,U_\vf\,.\label{S3}
\end{eqnarray}
As usual, $U_\rho$ and $U_\vf$ are the corresponding partial
derivatives of the potential $U$.

The key property of the dyonic configuration is that the
derivative of the potential,
\begin{equation}
U_\vf=\frac{a}{\rho}\left(P^2 e^{a\vf}-Q^2e^{-a\vf}\right)\,,
\end{equation} vanishes when
\begin{equation}
\vf=a^{-1}\ln\left|Q/P\right|\equiv \bvf\,.
\end{equation}
Therefore the constant dilaton $\vf=\bvf$ is a solution to the
dilaton equation of motion Eq.~(\ref{S3}), which is impossible for
a single-charged configuration. As a result, the dyonic dilaton
black hole admits Reissner-–Nordstr\"{o}m solution as a special
case.

Indeed, the equation for $\rho$ reads as:
\begin{equation}\label{S4}
\rho''=\frac{(D-3)\rho'^2}{(D-2)\rho}\,.
\end{equation}
It can be easily integrated, and the solution with Minkowski
asymptotic is $\rho=r^{D-2}$. Then from the constraint~(\ref{I7})
we find the remaining metric component:
\begin{equation}\label{S5}
    h=1-\frac{2M}{r^{D-3}}+\frac{2|QP|}{(D-3)(D-2)r^{2(D-3)}}\,,
\end{equation}
where the new integration constant $M$ appears.

The event horizon in Schwarzschild gauge emerges when $h$ is
vanishing. The equation~(\ref{S5}) for $h=0$ has two solutions,
$r=r_\pm$:
\begin{equation}\label{S5a}
    r_\pm=\left(\frac{M}{A_1\mp A_2}\right)^{1/(D-3)}\,,\quad\mbox{where}\quad  A_1\equiv \frac{(D-3)(D-2) M^2}{2|QP|},\quad A_2\equiv \sqrt{A_1^2- A_1}\,.
\end{equation}
With $A_{2}^2<0$ there would be a naked singularity, and with
$A_{2}^2=0$ the above solution describes an extremal
Reissner–-Nordstr\"{o}m black hole, when the two horizons
coincide. However, our study is focused on the non-extremal case,
when there are two distinct horizons, so in what follows we assume
that $A_{2}>0$.

\subsection{Dilaton `quantization' on RN background}
We see that static dilaton field provides Reissner–-Nordstr\"{o}m
metric. But is the solution stable if dilaton acquires some
dynamics? In order to check this, let us consider small deviations
of the dilaton field from constant solution: $\vf=\bvf+\dvf$,
where $\dvf\ll \bvf$. For simplicity, we will not carry out a full
stability analysis of the entire system, just trace the behavior
of solutions for $\dvf$ on a fixed RN background.

Then the dilaton equation~(\ref{S3}) with the given RN metric and
linearized with respect to the variable $\dvf$ will look like:
\begin{equation}\label{S6}
    \left(\left[(Mr^{3-D}-A_1)^2-A_2^2\right]r^{D-2}\dvf'\right)'=2 (D-3)^2 M^2 a^2
    r^{2-D}\dvf\,,
\end{equation}
where we used the formula $A_1 h=(Mr^{3-D}-A_1)^2-A_2^2$ following
from the Eqs.~(\ref{S5}--\ref{S5a}). After the coordinate change,
$x\equiv Mr^{3-D}-A_1$, the equation~(\ref{S6}) takes a notably
simple form:
\begin{equation}\label{S8}
    \frac{d}{dx}\left[\left(x^2-A_2^2\right)\frac{d}{dx}\dvf\right]=2a^2\dvf\,.
\end{equation}

In non-extremal case $A_{2}^2>0$, this is a Legendre differential
equation. Its non-singular solutions in the region $x^2\leq A_2^2$
between the two horizons are the Legendre polynomials $P_n$:
\begin{equation}\label{S9}
    \dvf=\dvfh P_n(A_2^{-1}x)\,,\quad n=0,1,2,...\,.
\end{equation}
Here $\dvfh$ is the integration constant: $\dvf$ is equal to $\pm
\dvfh$ on horizons located at $|x|=A_2$. From the Eq.~(\ref{S8})
we find that the non-negative integer parameter $n$ is related to
the dilaton coupling constant as following:
\begin{equation}\label{S10}
    a^2=n(n+1)/2,\quad n=0,1,2,...\,.
\end{equation}
Thus, here we encounter triangular numbers in the expression for
the dilaton coupling constant. Earlier this sequence arose when
considering series expansions and numerical solutions. Now, in
addition to previous approaches, we find a physical reason for the
appearance of triangular numbers: on a Reissner-–Nordstr\"{o}m
background the non-diverging solutions for $\dvf$ emerge only for
these discrete values of dilaton coupling constant.

Unlike the two horizons case, on the extremal RN background with
$A_2=0$, the non-singular solution to the Eq.~(\ref{S8}) exists
for any value of dilaton coupling constant:
\begin{equation}\label{S11}
    \dvf=C x^\nu, \quad\mbox{where}\quad
    a^2=\nu(\nu+1)/2\,,\quad C=\mathrm{const}\,.
\end{equation}
Obviously, the solution~(\ref{S11}) is analytic only in case of
triangular values of $a$. In~\cite{Galtsov:2014wxl} the same
result was derived by considering the series expansions at the
extremal horizon of the solution to the full
system~(\ref{S1}--\ref{S3}). Now we see that the non-analyticity
of dilaton solution for $\nu\neq n$ may be derived from the
dilaton behavior on RN background, both in the extremal and
non-extremal cases.

\section{The equivalent dynamical system}

The results obtained in the linear approximation can change when
the nonlinear dynamics is switched on. So now we consider the full
dyon-dilaton system described by the effective action~(\ref{I6}).
However, instead of solving a cumbersome system of
equations~(\ref{S1}--\ref{S3}), we transform the action~(\ref{I6})
to a more convenient form. It is easy to notice that its kinetic
part simplifies in the gauge $s=\rho$, in which horizons $h=0$ are
located at $r\ra\pm\infty$. Then the kinetic part is quadratic in
$h'\!/h$ and $\rho'\!/\rho$. It is convenient to choose
exponential variables for the metric:
\begin{equation}\label{I8}
    \sigma e^\chi=\frac{2(D-3)}{D-2}|QP|\,h\,,\quad
    \sigma e^\gamma=2(D-3)^2 h\rho^{2(D-3)/(D-2)}\,,\quad
    \sigma\equiv\mathrm{sign}(h)\,.
\end{equation}
In addition, we replace the dilaton function $\vf$ by its
deviation from constant solution:
\begin{equation}\label{I8a}
\psi=\vf-\bvf\,.
\end{equation}

With the new variables, the effective action reads as:
\begin{equation}
\frac{2(D-3)}{D-2}S_{\mathrm{eff}}=\int L_1(\gamma,\gamma')\,
dr-\int L_2(\chi,\chi',\psi,\psi')\,dr\,.\label{I9b}
\end{equation}
In this form it represents the sum of two independent actions. And
the latter describe just a one- and two-dimensional motion of
`particles' in the exponential potentials:
\begin{eqnarray}
&&L_1=\frac{\gamma'^2}{2}+\sigma e^\gamma\,,\label{I9a}\\
&&L_2=\frac{\chi'^2}{2}+\frac{\psi'^2}{2}+\sigma
\left(e^{\chi+a\psi}+e^{\chi-a\psi}\right)\,,\label{I9}
\end{eqnarray}
where the coordinate $r$ plays the role of `time'. It has long
been observed that spherically symmetric dyon-dilaton black hole
can be reformulated as a `Toda
molecule'~\cite{Gibbons:1987ps,Gibbons:1982ih}. Mention that black
hole's exterior region corresponds to $h>0$, and in subhorizon
region one has $h<0$. Therefore the
lagrangians~(\ref{I9a}--\ref{I9}) are actually different in the
two regions.

Obviously, the energy is conserved in both systems $L_1$ and
$L_2$. The conservation laws are:
\begin{eqnarray}
&&E_1=\frac{\gamma'^2}{2}-\sigma e^\gamma=\mathrm{const} \,,\\
&&E_2=\frac{\chi'^2}{2}+\frac{\psi'^2}{2}-\sigma
\left(e^{\chi+a\psi}+e^{\chi-a\psi}\right)=\mathrm{const}\,.\label{I9c}
\end{eqnarray}
However, the constants $E_1$ and $E_2$ are not independent: they
should obey the Hamiltonian constraint~(\ref{I7}) for the full
system:
\begin{equation}
H=E_1-E_2=0\,,\quad\Rightarrow\quad E_1=E_2\equiv E\,.
\end{equation}

The one-dimensional system given by the Eq.~(\ref{I9a}) can be
easily integrated:
\begin{eqnarray}
  &&h < 0\,,\; \mathrm{subhorizon}:\quad  e^\gamma=E\cosh^{-2}\left[\sqrt{E/2}(r-r_0)\right]\,,\label{1in}\\
  &&h > 0\,,\;\mathrm{exterior}:\quad
  e^\gamma=E\sinh^{-2}\left[\sqrt{E/2}(r-r_0)\right]\,.\label{1ext}
\end{eqnarray}
In subhorizon region $h<0$, the integration constant $r_0$ is not
important. The solution vanishes at $r\ra\pm\infty$, which
corresponds to the horizons $h=0$. In the exterior region $h>0$,
the outer horizon is located at $r\ra+\infty$ or $r\ra-\infty$,
where the solution approaches zero. The Minkowski asymptotic
corresponds to the vicinity of the boundary $r\ra r_0$. Indeed,
with $h\ra 1$, $r\ra r_0$ the line interval for
solution~(\ref{1ext}) can be written as:
\begin{equation}\label{ras}
    ds^2=-dt^2+d\tr^2+\tr^2 d\Omega^2_{D-2},\quad\mbox{where}\quad \tr\equiv
\left[(D-3)|r-r_0|\right]^{-1/(D-3)},
\end{equation}
which is indeed a Minkowski space.

The two-dimensional system~(\ref{I9}) seems to be non-integrable,
in general. Only two exact solutions with $a=1$ and $a=\sqrt{3}$
are known so far. In the first case the solution can be easily
found in terms of the variables $\chi\pm\psi$. It coincides up to
integration constants with the formulas for $\gamma$ given by the
Eqs.~(\ref{1in}--\ref{1ext}). The second known integrable case,
$a=\sqrt{3}$, is not so simple. The solution was found by reducing
the equations of motion to a system of equations for the
integrable Toda lattice~\cite{Gibbons:1982ih,Lee:1984hm}. In our
research, we want to put the question differently: instead of
establishing a correspondence between the equations of motion of
the dyon-dilaton system and the integrable Toda lattice for a
given dilaton coupling constant, we will determine at what values
of the coupling constant this correspondence exists.

\subsection{Toda-related solutions}

For a more general consideration, we consider the case when the
dilaton may interact differently with the gauge forms $F_{[2]}$
and $G_{[D-2]}$, as was discussed, for example,
in~\cite{Gibbons:1987ps}. The corresponding lagrangian can be
written as
\begin{equation}\label{TLag}
    L_2=\frac{\chi'^2}{2}+\frac{\psi'^2}{2}+\sigma
\left(e^{\chi+a_2\psi}+e^{\chi-a_1\psi}\right)\,,
\end{equation}
where there are two dilaton coupling constants $a_1$ and $a_2$.
They must be of the same sign, otherwise the system will not have
a constant dilaton solution, which is a distinctive feature of the
dyon configuration that we would like to preserve. The lagrangian
is invariant under the transformation $a_1\leftrightarrow a_2$,
$\psi\ra-\psi$. Therefore without loss of generality we can assume
that $a_1\geq a_2>0$.

In order to make the formulas simpler, we introduce a vector
$X=(\chi,\psi)$ and write the Hamiltonian for the
system~(\ref{TLag}):
\begin{equation}\label{Ham}
H_2=\frac12\sum_{i=1}^2 P_i^2-\sigma \sum_{i=1}^2 \exp\left(
    A_{ik}X^k\right)\,.
\end{equation}
Here
\begin{equation}\label{TM1}
P_i=\frac{dX^i}{dr}\,,\quad A_{ik}=\left(%
\begin{array}{cc}
  1 & a_2 \\
  1 & -a_1 \\
\end{array}%
\right)\,,
\end{equation}
and the expression $A_{ik}X^k$ implies the sum for $k=1,2$.
Exactly the same Hamiltonians in form were considered in the study
of generalized Toda lattices. Not all such systems are integrable.
However, a sufficient condition for integrability is well known.
In order to use it, we can calculate the following matrix:
\begin{equation}\label{TM2}
    C_{ik}=2\frac{\{A_{kl}X^l,\{A_{im}X^m,H_2\}\}}{\{A_{kl}X^l,\{A_{km}X^m,H_2\}\}}\,,
\end{equation}
where $\{\cdot,\cdot\}$ is a Poisson bracket. As was found in the
study of Toda lattices, if $C_{ik}$ is a Cartan matrix of some Lie
algebra, then the system~(\ref{Ham}) can be explicitly
integrated~\cite{Toda}.

For the Hamiltonian~(\ref{Ham}) we find that
\begin{equation}\label{Cartan}
C_{ik}=\left(%
\begin{array}{cc}
  2 & \frac{2(1-a_1a_2)}{1+a_1^2} \\
 \frac{2(1-a_1a_2)}{1+a_2^2} & 2 \\
\end{array}%
\right)\,.
\end{equation}
It is not difficult to calculate that $C_{ik}$ can be a Cartan
matrix only in four cases. The corresponding algebras and values
of dilaton couplings are listed below:
\begin{equation}\label{Cartan2}
\begin{tabular}{|c|c|c|c|}
\hline
 $A_1\oplus A_1$ & $A_2$ & $C_2$ & $G_2$ \\
  \hline
  $\left(%
\begin{array}{cc}
  2 & 0 \\
 0 & 2 \\
\end{array}%
\right)$ &  $\left(%
\begin{array}{cc}
  2 & -1 \\
 -1 & 2 \\
\end{array}%
\right)$ &  $\left(%
\begin{array}{cc}
  2 & -1 \\
 -2 & 2 \\
\end{array}%
\right)$ &  $\left(%
\begin{array}{cc}
  2 & -1 \\
 -3 & 2 \\
\end{array}%
\right)$ \\
  \hline
  $a_1a_2=1$ & $a_1=a_2=\sqrt{3}$ & $a_1=3,\,a_2=2$ &
  $a_1=3\sqrt{3},\,a_2=5/\sqrt{3}$\\
  \hline
\end{tabular}
\end{equation}

Exactly the same list of Lie algebras and dilaton coupling
constants (up to rescaling) was obtained in a series of papers by
Ivashchuk and
co-authors~\cite{Ivashchuk:2013jja,Abishev:2015pqa,Abishev:2017hmd}.
Note that, with the exception of the case $A_1\oplus A_1$, there
are only individual values corresponding to integrable solutions.
And when $a_1=a_2$ there are only two values: $1$ and $\sqrt{3}$.
Of course, we can not say that there are no other integrable
cases. However, there are no signs indicating the presence of a
triangular sequence $a^2=n(n+1)/2$, or any other sequence.

In the
works~\cite{Ivashchuk:2013jja,Abishev:2015pqa,Abishev:2017hmd},
the authors considered solutions only in the exterior region. And
we also need to obtain solutions in the subhorizon area, in order
to verify the regularity of both horizons. Moreover, in these
studies the authors did not construct explicit solutions for $C_2$
and $G_2$ cases. Therefore, now we partially reproduce, and
partially supplement their results, using the same approach that
was developed by Ivashchuk and co-authors
in~\cite{Ivashchuk:2002jr,Golubtsova:2008ua}.

The equations of motion for the lagrangian~(\ref{TLag}) are:
\begin{equation}\label{Teq1}
    \frac{d^2X^i}{dr^2}=\sigma \sum_{k=1}^2 (A^T)^{ik} \exp\left(
    A_{kl}X^l\right)\,,\quad i=1,2\,.
\end{equation}
First we need to find a linear transformation of variables $X=KY$
with a constant matrix $K_{ik}$, that reduces this system of
equations to the form
\begin{equation}\label{Teq2}
   \frac{d^2Y^i}{dr^2}=-\sum_{k=1}^2 B^{ik} \exp\left(
    C_{kl}Y^l\right)\,,\quad i=1,2\,,
\end{equation}
where $B^{ik}$ is some constant diagonal matrix. This system
coincides with open Toda lattice equations corresponding to the
Lie algebra with the Cartan matrix $C_{ik}$.

It is easy to find that
\begin{equation}\label{TM3}
    K=A^{-1}C=\left(%
\begin{array}{cc}
  \frac{2}{1+a_2^2} & \frac{2}{1+a_1^2} \\
  \frac{2a_2}{1+a_2^2} & -\frac{2a_1}{1+a_1^2} \\
\end{array}%
\right),\; B=-\sigma
C^{-1}\!AA^T=-\frac{\sigma}{2}\,\mathrm{diag}\left(1+a_2^2,1+a_1^2\right)\,.
\end{equation}
An important parameters of the solution are components of the
twice dual Weyl vector in the basis of simple roots of the
corresponding Lie Algebra:
\begin{equation}\label{Tn}
    n_i=2\sum_{k=1}^2 \left(C^{-1}\right)^{ik}\,.
\end{equation}
For the Cartan matrix~(\ref{Cartan}) we have
\begin{equation}\label{Tn2}
    n=\left(%
\begin{array}{cc}
  \frac{a_1(1+a_2^2)}{a_1+a_2}\,, & \frac{a_2(1+a_1^2)}{a_1+a_2} \\
\end{array}%
\right)\,,\quad
    \begin{tabular}{|c|c|c|c|c|}
      \hline
      & $A_1\oplus A_1$ & $A_2$ & $C_2$ & $G_2$ \\
      \hline
      $n_1$ & 1 & 2 & 3 & 6\\
      $n_2$ & 1 & 2 & 4 & 10\\
      \hline
    \end{tabular}
\end{equation}

The key part of the solution are the so-called fluxbrane
polynomials:
\begin{equation}\label{TH}
    H_i(z)=c_i+\sum_{k=1}^{n_i}p_{ik}z^k(r)\,,\quad i=1,2\,,\quad \mbox{where}\quad
    z\equiv-\frac{2\sigma}{E} e^{-\sqrt{2E} (r-r_0)}\,.
\end{equation}
The four constants $c_i>0$, $E>0$ and $r_0$ are the free
parameters of solution which can be found from initial or boundary
conditions. Mention that $E$ and $r_0$ are common parameter for
systems $L_1$ and $L_2$. The ansatz for the system~(\ref{Teq2})
reads as
\begin{equation}\label{TsolY}
    Y^i=-n_i\sqrt{E/2}(r-r_0)-\ln H_i(z)\,,\quad i=1,2\,.
\end{equation}
Substituting it into the equations, we obtain the values of the
parameters $p_{ik}$ for which the ansatz satisfies equations. The
resulting fluxbrane polynomials are given in the Appendix by the
Eqs~(\ref{A1}--\ref{G22}).

All coefficients of the fluxbrane
polynomials~(\ref{A1}--\ref{G22}) are positive when $c_1,c_2>0$.
They depend on the variable $z$ which is non-negative in
subhorizon region. As a result, the functions $\ln H_i(z)$ are
well-defined everywhere on the interval between the two horizons.
In the exterior region $z\leq 0$, and polynomials $H_i(z)$ can
take negative values. However, $H_i(0)>0$ which corresponds to the
horizon $r\ra +\infty$. Therefore there exist such interval
$r_s<r<+\infty$, where $H_i(z)>0$. Recall that in the chosen gauge
$s=\rho$ the exterior region (from outer horizon to Minkowski
space) corresponds to the interval $r_0\leq r<+\infty$, as it
follows from the Eq.~(\ref{1ext}). With a suitable choice of
parameters $r_0$ and $c_i$ we can always ensure that $r_s<r_0$,
and the solutions for $Y^i$ are well-defined in the exterior
region also.

It is interesting that by construction the asymptotics of $Y^i$
are similar on both horizons:
\begin{equation}\label{TYas}
    Y^i|_{r\ra\pm\infty}=-n_i\sqrt{E/2} |r|\,.
\end{equation}
Now let us turn to the original functions, $\chi$ and $\psi$.
Calculating $X=KY$ we find:
\begin{eqnarray}\label{TsolX}
\chi&=&\frac{1}{a_1+a_2}\sum_{i=1}^2 a_i\ln\left(\frac{-\sigma
E}{2}z
H_i^{-\frac{2}{n_i}}(z)\right)\,,\\
\psi&=&\frac{2a_1a_2}{a_1+a_2}\ln\left(H_1^{-\frac{1}{n_1}}\!(z)\,
H_2^{\frac{1}{n_2}}(z)\right)\,.
\end{eqnarray}
As it should be, the function $\chi$ goes to $-\infty$ on both
horizons:
\begin{equation}\label{Thas}
    \chi|_{r\ra\pm\infty}=-\sqrt{2E}|r|\,.
\end{equation}
The dilaton function on both horizons approaches constants:
\begin{eqnarray}
  r &\ra& -\infty:\quad \psi \ra \psi_{\mathrm{in.\,hor.}}= \frac{2a_1a_2}{a_1+a_2}\ln\left(p_{1n_1}^{-\frac{1}{n_1}}p_{2n_2}^{\frac{1}{n_2}}\right)\,,\\
  r &\ra& +\infty:\quad \psi \ra \psi_{\mathrm{out.\,hor.}}=
  \frac{2a_1a_2}{a_1+a_2}\ln\left(c_{1}^{-\frac{1}{n_1}}c_{2}^{\frac{1}{n_2}}\right)\,.
\end{eqnarray}
Thus, analytical solutions turn out to be regular on both horizons
and outside the black hole.

It is worth mentioning that for $a_1=a_2$ the lagrangian $L_2$
becomes invariant with respect to the reflection $\vf\ra -\vf$. As
a result, the dilaton function $\psi$ assumes the same modulus
values on the horizons. For the expressions for the fluxbrane
polynomials~(\ref{A1}--\ref{A2}) given in the Appendix we have in
$A_1\otimes A_1$ case:
$\psi_{\mathrm{out.\,hor.}}=-\psi_{\mathrm{in.\,hor.}}=\ln(c_2/c_1)$;
in $A_2$ case:
$\psi_{\mathrm{in.\,hor.}}=\psi_{\mathrm{out.\,hor.}}=\ln(c_2/c_1)$.

The Minkowski asymptotic is achieved at $\sigma=1$, $r=r_0$, which
implies $z_0=-2/E$. Then the condition $h(r_0)=1$ imposes an
additional constraint
\begin{equation}\label{Thas2}
   H_1^{-\frac{2}{1+a_2^2}}(z_0)H_2^{-\frac{2}{1+a_1^2}}(z_0)=\frac{2(D-3)}{D-2}|QP|
\end{equation}
on the parameters $c_i$ and $E$. Next, the asymptotic for the
metric $h=1-2M\tilde{r}^{3-D}+\ldots$, where $\tilde{r}$ is given
by the Eq.~(\ref{ras}), provides the relation between the mass
parameter $M$ and $c_i$, $E$:
\begin{equation}\label{TMass}
    M=\frac{1}{(D-3)(a_1+a_2)}\sqrt{\frac{E}{2}}\sum_{i=1}^2
    a_i\left(1+\frac{4}{n_i E}\left.\frac{d\ln
    H_i(z)}{dz}\right|_{z=z_0}\right)\,.
\end{equation}
Expanding dilaton function $\vf$ as $\vf=\vf_\infty+\Sigma\,
\tilde{r}^{3-D}+\ldots$, we can find that
\begin{equation}\label{Tpsias}
    \vf_\infty=\bvf+\psi(z_0)\,,\quad
    \Sigma=\frac{2}{D-3}\sqrt{\frac{2}{E}}\left.\frac{d\psi}{dz}\right|_{z=z_0}\,.
\end{equation}
Thus, there are four equations~(\ref{Thas2}--\ref{Tpsias}) that
allow us to express three parameters $c_i$, $E$ in terms of four
parameters $\vf_\infty,\, M\,,|QP|\,,\Sigma$. This means that
those four parameters are not independent, as expected.

It is also possible to calculate the Hawking temperature of the
black holes. We give the computations for the case $D=4$, to avoid
cumbersome factors that depend on the dimension of the spacetime:
\begin{equation}\label{TT}
    T_H=\frac{1}{2\pi}\left(\sqrt{g^{rr}}\frac{\p\sqrt{-g_{tt}}}{\p r}\right)_{r\ra+\infty}=\frac{1}{2\pi
    }\left(e^{\chi-\gamma}\chi'\right)_{r\ra+\infty}=\frac{1}{4\pi\sqrt{2E}}\,c_1^{-\frac{2}{1+a_2^2}}c_2^{-\frac{2}{1+a_1^2}}\,.
\end{equation}
In case of single dilaton constant, $a_1=a_2=a$, one has $T=
(4\pi\sqrt{2E})^{-1}\exp(a^{-1}\psi_{\mathrm{out.\,hor.}})$.

Summarizing, we can see that the configurations with one coupling
constant, $a=1$ and $a=\sqrt{3}$, are the only ones that can be
easily associated with the generalized Toda lattices related to
Lie algebras. Therefore, there is no reason to extend the
properties of these solutions to cases of other dilaton coupling
values. In the next section, we show that, in fact, solutions with
different values of the parameter $a$ have essentially different
properties.

\section{Numerical solutions}

The full dyon-dilaton system~(\ref{I6}) is most likely
non-integrable for arbitrary values of dilaton coupling constant.
However, it can be divided into two independent
subsystems~(\ref{I9a}) and~(\ref{I9}). The first one is explicitly
integrated, so we need to solve numerically only the second one.
This section is devoted to the construction of numerical solutions
for the dynamical system given by the lagrangian $L_2$. Our goal
is to build the regular solutions, i.e. for which the values of
the metric and dilaton functions and their derivatives are finite
on horizons. Therefore we will only consider a subhorizon region
with $h\leq 0$. This means that we choose $\sigma=-1$ in the
expression~(\ref{I9}) for $L_2$.

In previous section we considered the connection between the
system $L_2$ and the Toda molecule, but now it is more convenient
to treat $L_2$ as describing a particle moving in the
$(\chi,\psi)$-plane in the potential
$e^{\chi-a\psi}+e^{\chi+a\psi}$. This allows to predict in advance
the basic properties of solutions. The free asymptotic motion at
$r\ra\pm\infty$ happens when potential vanishes. The potential
goes to zero if $\chi\ra -\infty$ and $a|\psi|<|\chi|$. So,
asymptotically the particle moves with constant velocity
$\chi',\psi'=\mathrm{const}$, and the direction of the velocity
vector can be any, provided that $a|\psi'|<|\chi'|$. The regular
horizon in the dyon-dilaton system implies $\chi\ra-\infty$,
$\psi\equiv\psih=\mathrm{const}$. It corresponds to an asymptotic
motion with a velocity vector parallel to the $\chi$-axis. We can
choose the initial condition in such a way as to satisfy the
requirement of regularity of one of the horizons:
$\psi'|_{r\ra-\infty}=0$ and $\psi|_{r\ra-\infty}=\psih$. However,
in the general case, the asymptotics near the second horizon can
be different: the velocity vector can be unparallel to the
$\chi$-axis, as $r\ra+\infty$. Only potentials possessing a hidden
internal symmetry can guarantee a transition from one nonsingular
asymptotic to another. We examined such symmetries in the previous
section and made sure that they are rather rare.

The potential $e^{\chi-a\psi}+e^{\chi+a\psi}$ has form of a
symmetric valley. Thus the particle trajectory from $r\ra-\infty$
to $r\ra+\infty$ oscillates in $\psi$-direction and crosses the
$\chi$-axis for several times. The number of intersections of the
trajectory with the $\chi$-axis divides the set of solutions into
families, each of which is characterized by its number of
intersections, $n$. The solutions which separate families with $n$
and $(n+1)$ intersections are separatrixes. They separate the
solutions with different asymptotics at $r\ra+\infty$, because
when $n$ is odd, the signs of the asymptotic values
$\psi|_{r\ra+\infty}$, $\psi'|_{r\ra+\infty}$ are opposite to the
sign of the initial value, $\psih$; when $n$ is even, the signs
are the same. The separatrix beginning at $(-\infty,\psih)$ will
end at $(-\infty,(-1)^n\psih)$ due to symmetry of the lagrangian
$L_2$ with respect to reflection $\psi\ra-\psi$ and time inversion
$r\ra -r$. Therefore the separatrix solutions have finite
$\psi$-coordinate everywhere and correspond to the dyon-dilaton
black holes with two regular horizons. Our goal is to construct
such solutions numerically and find what values of the system
parameters they correspond to.

The parameter $E$ can be set equal to unity by the rescaling
$\chi\ra\chi+\ln E$, $r\ra r/\sqrt{E}$. Thus, we have two
parameters that determine the behavior of the solution: $\psih$
and $a$. For the convenience of testing the triangular hypothesis,
we will use the parameter $\nu$ instead of $a$. They are related
as $\nu(\nu+1)/2=a^2$. According to the hypothesis, separatrix
solutions must exist only for integer values of the parameter
$\nu$.

\begin{figure}[tb]
\hbox to\linewidth{\hss%
\psfrag{a}{{$\nu$}} \psfrag{b}{\Large{$\psih$}}
\includegraphics[width=0.75\linewidth,height=0.45\linewidth]{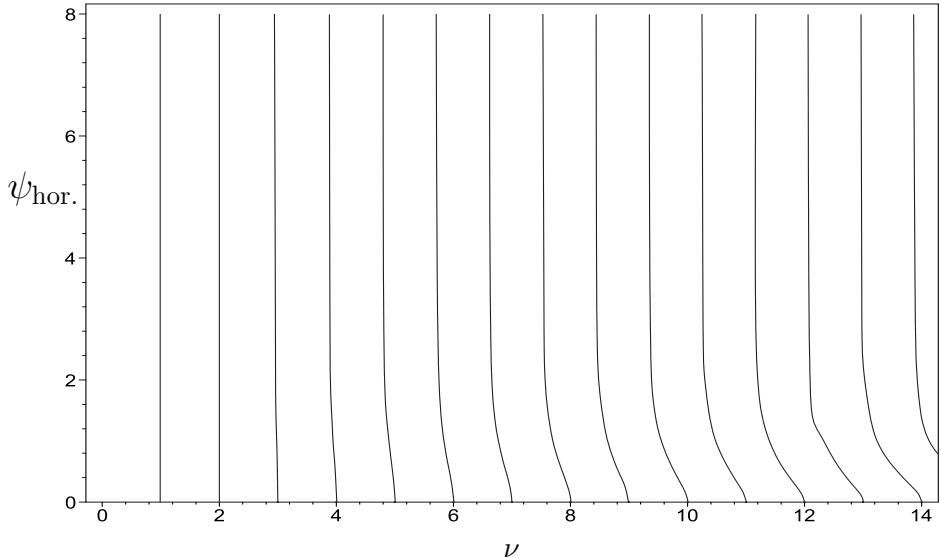}
\hss} \caption{\small   The curves $\psih(\nu)$ corresponding to
non-extremal regular black hole solutions. The regions between the
curves are domains of solutions singular on one of the horizons.}
\label{F1}
\end{figure}

On the Fig.~\ref{F1} we have depicted the sets of points on
$(\nu,\psih)$-plane to which the numerically constructed
separatrix solutions correspond. They form a set of lines, each of
which corresponds to a family of solutions with $n=1,2,3,\ldots$
intersections of the $\chi$-axis. However, these families can be
combined into two classes.
\begin{itemize}
    \item The first class contains the two vertical lines $\nu=1$
    and $\nu=2$. The corresponding solutions are the known
    integrable ones (with $a=1,\sqrt{3}$). They correspond to discrete
    values of the dilaton coupling constant, while the parameter
    $\psih$ can be arbitrary.
    \item The lines belonging to the second class are the curves that,
    for small values of the parameter $\psih$, are close to vertical lines $\nu=3,4,5,\ldots$,
    but with an increase in the parameter, they are noticeably deviated from
    them. Thus, regular solutions for $\nu>2$ can be found in the
    intervals $\nu_{n}^{\mathrm{min}}\leq\nu<n$, where $n$ is a
    number of intersections with the $\chi$-axis for a given family of
    solutions. For $\psih\ra 0$ the curves approach $\nu=n$ straights, which corresponds to the solutions obtained in the linear
approximation for small values of the dilaton. However, for large
$\psih$ the non-linearity deforms the curves, which then tend to
straights $\nu=\nu_{n}^{\mathrm{min}}<n$. For relatively small
values of dilaton coupling constant $(\nu\leq 6)$, the intervals
$(\nu_{n}^{\mathrm{min}},n)$ are very narrow. Apparently,
therefore, it was conjectured that they reduce to discrete values
$\nu=n$. For large $\nu$, the intervals become quite wide.

The qualitative behavior of the solutions depends not only on the
parameter $\nu$, but also on $\psih$. For
    given $\nu$ there is only a discrete set of values
    $\{\psih^{1},..,\psih^{k}\}$, $k=k(\nu)=0,1,2,\dots$, which corresponds to regular solutions.
    This set can be empty for $\nu \lesssim 12.07$, because there are
    gaps between the intervals $(\nu_{n}^{\mathrm{min}},\nu=n)$, where there are no regular solutions.
    But for $\nu \gtrsim 12.07$ the intervals completely cover the
    $\nu$-axis and even start overlapping each other for $\nu\gtrsim 12.96$, so there can be one or two distinct regular solutions for given $\nu$.
In the case of overlap, the solutions differ in the number of
zeros of the dilaton function, which can be equal to $[\nu+1]$ or
$[\nu+2]$, where $[\cdot]$ denotes the integer part of a number.
For $\nu\gtrsim 30$ the
    overlap can be double (resulting in three distinct solutions with
    $[\nu+1]$,
$[\nu+2]$ and $[\nu+3]$ zeroes of the dilaton functions), e.t.c.
Probably, for $\nu\ra\infty$ the
    overlap can be infinite, however, it is difficult to verify by numerical
    calculations.
\end{itemize}

We can formulate the result of our numerical study as follows: the
spectrum of dilaton coupling constant corresponding to regular
solutions is discrete for $n\leq 2$, continuous with gaps for
$2<\nu\lesssim 12$, and continuous for $\nu \gtrsim 12.07$. This
disproves the triangular hypothesis that the spectrum should only
be discrete. However, now numerical results allow us to put
forward the hypothesis that the spectrum of a parameter $\psih$
must be discrete, with the exception of two special cases with
$a=1,\sqrt{3}$, for which the spectrum is continuous.

It is important to mention that the authors of the triangular
hypothesis themselves wrote in their work~\cite{Poletti:1995yq}
that ``the precise numerical value of $a$ begins to show some
dependence on the initial data for large values of $\psih$'' (they
actually used another notations for dilaton coupling constant and
dilaton value at horizon). They found that ``even for large $a$
the largest values of $\psih$ consistent with asymptotically flat
solutions yield numerical values of $a$ which agree with the
[triangular] series to 0.1\%. We have numerically checked terms up
to $a=\sqrt{21}$ in the series''. Then they concluded that ``it
would appear to be merely the result of numerical errors''.
However, in present work, we checked the sequence up to
$a=\sqrt{465}$ and for much larger values of $\psih$. Look, for
example, on the Fig.~\ref{F1} where we present the graphics for
$a\leq\sqrt{105}$. It is easy to see that for the range
$a\leq\sqrt{21}$ ($\nu\leq 6$) considered
in~\cite{Poletti:1995yq}, deviations from the triangular sequence
are really small. However, then they become more and more
significant, and they can not be attributed in any way to the
errors of numerical integration.

By the way, our choice of the gauge $s=\rho$ is especially
convenient for numerical integration, since the most problematic
areas for numerical integration are the areas near the horizons.
And in this gauge they are stretched to infinite intervals, which
greatly simplifies the verification of the convergence of the
numerical scheme. For example, for each numerical solution we
checked the feasibility of the energy constraint~(\ref{I9c}). If
the solutions contained errors, they would not satisfy it.
However, for all solutions the constraint was satisfied with
sufficient accuracy. We used the standard MAPLE tools for the
numerical integration of differential equations with an absolute
and relative errors of not more than $10^{-13}$.

Let us make some more remarks. Consider the transition to extremal
black hole solution, i.e. when the distance between the horizons
goes to zero. For every solution the number of intersection with
$\chi$-axis (zeroes of the dilaton function) does not depend on
the distance between the horizons. As a result, the dilaton
function will oscillate rapidly, with a `period' of oscillations
tending to zero. Its contribution into energy density,
$\psi'^2\!/2$, will diverge if the amplitude of the oscillations
associated with the parameter $\psih$ does not tend to zero.
However, for a given value of the dilaton coupling constant, the
parameter $\psih$ can be set to zero, provided that the regularity
of solutions is maintained, for only two cases: $a=1,\sqrt{3}$.
Otherwise, the spectrum of $\psih$ is discrete, and the transition
to an extreme black hole is singular.

\section{Conclusion}

We studied static spherically-symmetric non-extremal black hole
solutions in the dyon-dilaton system, and found that the condition
for the existence of regular solutions (non-singular on both
horizons and with Minkowski asymptotic) leads to discreteness in
the parameter space of the model. However, the entire set of
regular solutions is divided into two essentially different
classes.

The first class includes solutions arising in configurations with
a few special values of the dilaton coupling constant(s), which
allow one to establish a connection with solutions for Toda
lattices. Even in the presence of two dilaton coupling constants
there are only four families of solutions in this class that
correspond to Lie algebras $A_1\oplus A_1$, $A_2$, $C_2$ and
$G_2$. From this point of view, there is absolutely no
confirmation of the triangular hypothesis.

For the existence of regular solutions belonging to the second
class, it is not required that the dilaton coupling constant takes
on any special values. At small values of $a$, there are intervals
in which there are no regular solutions, but these `voids'
disappear after $a\gtrsim 8.9$. In this class there is,
apparently, a countable number of families of regular solutions.
However, in contrast to solutions of the first class, solutions
from the second class are separatrixes that separate singular
solutions with different numbers of zeros of the dilaton function.
For a given value of the dilaton coupling constant, there can be
several regular solutions differing by the number of zeros.
Without changing the dilaton coupling constant, it is impossible
to go from one regular solution to another with a continuous
change in the parameters of the system.

This property resembles a kind of quantization for dyonic dilaton
black holes. Indeed, the dilaton coupling constant is an external
parameter in the model, in the real world it is most probably
fixed and its value is determined by some fundamental theory. If
its value is large enough, for example, $a=9.52$, then there may
exist regular dyonic black holes with 12 and 13 zeroes of dilaton
function in the subhorizon region. A nontrivial question arises:
how can a transition from one state to another occur if they are
separated by singular configurations? Of course, the
non-stationary processes go far beyond our study of static
solutions. Nonetheless, the process of transition between regular
solutions, separated by singular states, can be accompanied by
intriguing physical phenomena. Taking into account that the
dyon-dilaton models are used to describe condensed matter
systems~\cite{Kundu:2012jn,Amoretti:2016cad}, configurations with
regular solutions separated by singular states, similar to those
studied in this paper, can find application in this field.

The discreteness in the parameter space corresponding to regular
solutions is also observed with various modifications of the
model, such as adding a cosmological
constant~\cite{Poletti:1995yq} or higher-curvature
corrections~\cite{Chen:2008hk}. Indeed, as we learned from the
verification of the triangular hypothesis, the discreteness arises
when considering the dilaton on the Reissner–-Nordstr\"{o}m
background, and the smooth inclusion of extra terms will
inevitably coexist with the discreteness, at least to some extent.

A possible topic for further research is also the study of
rotating dyonic dilaton black
holes~\cite{Clement:1986bt,Rasheed:1995zv}. They are also
characterized by the presence of a special value of the dilaton
coupling constant, which is $a=\sqrt{3}$~\cite{Galtsov:1995mb}.
One can expect that for rotating solutions the discreteness in the
parameter space also manifests itself in a nontrivial fashion. It
would be especially interesting to explore the system of dyonic
diholes~\cite{Cabrera-Munguia:2015fha,Garcia-Compean:2015ywa,Clement:2017otx}
in presence of a dilaton field.

This work was financially supported by the Ministry of Education
and Science of the Russian Federation (the Agreement number
02.a03.21.0008) and by the Russian Foundation for Fundamental
Research under grant 17-02-01299.

\section*{Appendix: Fluxbrane polynomials}
Here we give fluxbrane polynomials that are part of the
solutions~(\ref{TsolY}) of the open Toda lattice
equations~(\ref{Teq2}) corresponding to the Lie algebras arising
in the models of dyon-dilaton black holes.
\begin{description}
    \item[Case $A_1 \oplus A_1$:]
    \begin{equation}\label{A1}
    H_1=c_1+\frac{(1+a_2^2)}{8c_1}z\,,\quad H_2=c_2+\frac{(1+a_1^2)}{8c_2
    }z\,,\quad \mbox{where}\quad a_1a_2=1\,.
\end{equation}
    \item[Case $A_2$:]
    \begin{equation}\label{A2}
    H_1=c_1+\frac{c_2}{2c_1}z+\frac {1}{
16c_2}z^2\,,\quad H_2=c_2+\frac{c_1}{2c_2}z+\frac {1}{
16c_1}z^2\,.
\end{equation}
    \item[Case $C_2$:]
    \begin{eqnarray}
      && H_1=c_1+\frac{5c_2}{8c_1}z+\frac{5^2c_1}{2^7 c_2}z^2+\frac{5^3}{2^{10}3^2 c_1}z^3\,,\label{C21}\\
      &&
      H_2=c_2+\frac{5c_1^2}{4c_2}z+\frac{5^2}{2^6}z^2+\frac{5^3c_2}{2^8 3^2c_1^2}z^3+\frac{5^4}{2^{14} 3^2 c_2}z^4\,.\label{C22}
    \end{eqnarray}
    \item[Case $G_2$:]
    \begin{eqnarray}
      && H_1=c_1+\frac{7c_2}{6c_1}z+\frac{7^2c_1^2}{48c_2}z^2+\frac{7^3}{2^4 3^4}z^3+\frac{7^4 c_2}{2^8 3^5  c_1^2}z^5+\frac{7^5 c_1}{2^9 3^5 5^2c_2}z^5+\frac{7^6}{2^{12}3^7 5^2 c_1}z^6\,,\label{G21}\\
      && H_2=c_2+\frac{7c_1^3}{2c_2}z+\frac{7^2c_1}{16}z^2\!+
      \frac{7^3c_2}{2^3 3^3c_1}z^3\!+\frac{7^4(c_1^5+c_2^3)}{2^8 3^3c_1^3
      c_2}z^4\!+\frac{7^6}{2^8 3^4 5^2}z^5\!+
      \frac{7^6(3^3c_1^5+5^2c_2^3)}{2^{12}3^7 5^2 c_1^2
      c_2^2}z^6\!+\nn\\
      &&
      \quad\quad\frac{7^7c_1}{2^{11} 3^7 5^2c_2}z^7+
      \frac{7^8}{2^{16} 3^8 5^2c_1}z^8+
      \frac{7^9 c_2}{2^{17} 3^{12}5^2c_1^3}z^9+
      \frac{7^{10}}{2^{20} 3^{12} 5^4 c_2}z^{10}\,.\label{G22}
    \end{eqnarray}
\end{description}


\begin{thebibliography}{99}

\bibitem{Gibbons:1987ps}
  G.~W.~Gibbons and K.~i.~Maeda,
  {\it Nucl.\ Phys.\ B} {\bf 298} (1988) 741.

\bibitem{Garfinkle:1990qj}
  D.~Garfinkle, G.~T.~Horowitz and A.~Strominger,
{\it  Phys.\ Rev.\ D} {\bf 43} (1991) 3140.
   Erratum: [Phys.\ Rev.\ D {\bf 45} (1992) 3888].

\bibitem{Dobiasch:1981vh}
  P.~Dobiasch and D.~Maison,
{\it  Gen.\ Rel.\ Grav.}  {\bf 14} (1982) 231.

\bibitem{Gibbons:1982ih}
  G.~W.~Gibbons,
 {\it Nucl.\ Phys.\ B} {\bf 207} (1982) 337.

\bibitem{Lee:1984hm}
  S.~C.~Lee,
 {\it Phys.\ Lett.}  {\bf 149B} (1984) 98.

\bibitem{Poletti:1995yq}
  S.~J.~Poletti, J.~Twamley and D.~L.~Wiltshire,
  {\it Class.\ Quant.\ Grav.}  {\bf 12} (1995) 1753,
   [Erratum: {\it Class.\ Quant.\ Grav.}  {\bf 12} (1995) 2355]

\bibitem{Gibbons:1993xt}
  G.~W.~Gibbons, D.~Kastor, L.~A.~J.~London, P.~K.~Townsend and J.~H.~Traschen,
 {\it Nucl.\ Phys.\ B} {\bf 416} (1994) 850.

\bibitem{Nozawa:2010rf}
  M.~Nozawa,
 {\it Class.\ Quant.\ Grav.}  {\bf 28} (2011) 175013.

\bibitem{Galtsov:2014wxl}
  D.~Gal'tsov, M.~Khramtsov and D.~Orlov,
  {\it Phys.\ Lett.\ B} {\bf 743} (2015) 87.

\bibitem{Toda}
B.~Constant, {\it Adv.\ Math.} {\bf 34} (1979) 195.

\bibitem{Ivashchuk:2013jja}
  V.~D.~Ivashchuk,
  {\it J.\ Geom.\ Phys.}  {\bf 86} (2014) 101.

\bibitem{Abishev:2015pqa}
  M.~E.~Abishev, K.~A.~Boshkayev, V.~D.~Dzhunushaliev and V.~D.~Ivashchuk,
{\it  Class.\ Quant.\ Grav.}  {\bf 32} (2015) no.16,  165010.

\bibitem{Abishev:2017hmd}
  M.~E.~Abishev, K.~A.~Boshkayev and V.~D.~Ivashchuk,
{\it  Eur.\ Phys.\ J.\ C} {\bf 77} (2017) no.3,  180.

\bibitem{Ivashchuk:2002jr}
  V.~D.~Ivashchuk,
  {\it Class.\ Quant.\ Grav.}  {\bf 19} (2002) 3033.

\bibitem{Golubtsova:2008ua}
  A.~A.~Golubtsova and V.~D.~Ivashchuk,
  arXiv:0804.0757 [nlin.SI].

\bibitem{Kundu:2012jn}
  N.~Kundu, P.~Narayan, N.~Sircar and S.~P.~Trivedi,
 {\it JHEP} {\bf 1303} (2013) 155.

\bibitem{Amoretti:2016cad}
  A.~Amoretti, M.~Baggioli, N.~Magnoli and D.~Musso,
  {\it JHEP} {\bf 1606} (2016) 113.

\bibitem{Chen:2008hk}
  C.~M.~Chen, D.~V.~Gal'tsov and D.~G.~Orlov,
{\it  Phys.\ Rev.\ D} {\bf 78} (2008) 104013.

\bibitem{Clement:1986bt}
  G.~Cl\'{e}ment,
{\it  Phys.\ Lett.\ A} {\bf 118} (1986) 11.

\bibitem{Rasheed:1995zv}
  D.~Rasheed,
 {\it Nucl.\ Phys.\ B} {\bf 454} (1995) 379.

\bibitem{Galtsov:1995mb}
  D.~V.~Galtsov, A.~A.~Garcia and O.~V.~Kechkin,
 {\it Class.\ Quant.\ Grav. }  {\bf 12} (1995) 2887.

\bibitem{Cabrera-Munguia:2015fha}
  I.~Cabrera-Munguia, C.~Lämmerzahl and A.~Macías,
 {\it Phys.\ Lett.\ B} {\bf 743} (2015) 357.

\bibitem{Garcia-Compean:2015ywa}
  H.~Garc\'{i}a-Compe\'{a}n and V.~S.~Manko,
 {\it Phys.\ Lett.\ B} {\bf 748} (2015) 366.

\bibitem{Clement:2017otx}
  G.~Cl\'{e}ment and D.~Gal'tsov,
{\it  Phys.\ Lett.\ B} {\bf 773} (2017) 290.

\end{thebibliography}
\end{document}